\documentclass[lettersize,journal]{IEEEtran}
\usepackage{amsmath,amsfonts}
\usepackage{algorithmic}
\usepackage{array}
\usepackage[caption=false,font=normalsize,labelfont=sf,textfont=sf]{subfig}
\usepackage{textcomp}
\usepackage{stfloats}
\usepackage{url}
\usepackage{verbatim}
\usepackage{graphicx}
\usepackage{xcolor}

\def\BibTeX{{\rm B\kern-.05em{\sc i\kern-.025em b}\kern-.08em
    T\kern-.1667em\lower.7ex\hbox{E}\kern-.125emX}}
\usepackage{balance}
\begin{document}

\title{Low-loss Si-based Dielectrics for High Frequency Components of Superconducting Detectors}

\author{%
M. Lisovenko, Z. Pan, P. S. Barry, T. Cecil, C. L. Chang, R. Gualtieri, J. Li, V. Novosad, G. Wang, and V. Yefremenko

\thanks{
(\textit{Corresponding author: Marharyta Lisovenko)}}
\thanks{M. Lisovenko is with the Argonne National Laboratory, 9700 South Cass Avenue, Lemont, IL, 60439, USA (email: mlisovenko@anl.gov)}%
\thanks{P. Barry is with Cardiff University, Cardiff CF10 3AT, UK (email: barryp2@cardiff.ac.uk)}
\thanks{Z. Pan, T. Cecil, R. Gualtieri, J. Li, V. Novosad, G. Wang, and V. Yefremenko are with the Argonne National Laboratory, 9700 South Cass Avenue, Lemont, IL, 60439, USA (emails: zpan@anl.gov; cecil@anl.gov; rgualtieri@anl.gov; juliang.li@anl.gov; novosad@anl.gov; gwang@anl.gov; yefremenko@anl.gov)}%
\thanks{C. L. Chang is with the Argonne National Laboratory, Argonne, IL 60439 USA, and the University of Chicago, 5640 South Ellis Avenue, Chicago, IL 60637 USA (email: clchang@kicp.uchicago.edu) }%

}

\markboth{Journal of \LaTeX\ Class Files,~Vol.~0, No.~0, April~2023}%
{How to Use the IEEEtran \LaTeX \ Templates}

\maketitle

\begin{abstract}
Silicon-based dielectric is crucial for many superconducting devices, including high-frequency transmission lines, filters, and resonators. Defects and contaminants in the amorphous dielectric and at the interfaces between the dielectric and metal layers can cause microwave losses and degrade device performance. Optimization of the dielectric fabrication, device structure, and surface morphology can help mitigate this problem. We present the fabrication of silicon oxide and nitride thin film dielectrics. We then characterized them using Scanning Electron Microscopy, Atomic Force Microscopy, and spectrophotometry techniques. The samples were synthesized using various deposition methods, including Plasma-Enhanced Chemical Vapor Deposition and magnetron sputtering. The film’s morphology and structure were modified by adjusting the deposition pressure and gas flow. The resulting films were used in superconducting resonant systems consisting of planar inductors and capacitors. Measurements of the resonator properties, including their quality factor, were performed. 
\end{abstract}


\section{Introduction}
High-quality dielectrics are crucial materials for a wide range of applications in superconducting devices. Silicon-based materials, such as amorphous silicon and carbide-, oxide- and nitride-compounds are of interest, and various deposition techniques can achieve low-loss mm-wave device dielectrics. These films can be fabricated by physical or chemical vapor deposition, and evaporation, combined with different modifications such as plasma-enhanced chemical vapor deposition (PECVD) \cite{Buijtendorp2020, Paik2010}, radio-frequency sputtering deposition \cite{Xu2003}, pulsed magnetron sputtering \cite{Tang2017}, and ion-beam assisted sputtering (IBAS) \cite{Cerny1998}. Each method has its advantages and drawbacks. For example, the widely used magnetron sputtering (MS) proved to be a promising approach for fabricating many devices with relatively low losses and applied temperatures in the range of 200-400~$^{\circ}$C. This method improves film quality by increasing atom mobility with RF bias and higher temperature on the substrate during film growth \cite{Freer1990}. However, high temperatures can degrade the characteristics of other microwave circuit components and exclude the use of the lift-off technique. We need to reduce the dielectric loss from two-level systems without complicating fabrication steps with high temperatures, which can cause topical problems in superconducting detector development \cite{Chang2015, Li2013, oconnel2008}.  

For this purpose, we consider two enhanced approaches of IBAS and PECVD. Ion-beam-assisted sputtering utilizes an ion source, which provides additional energy on the outermost growing atomic layer and allows control over film density and microstructure at room temperature. The same time the flux and the energy of ions can be changed independently of the flux of the depositing atoms, thus allowing easy control of the stoichiometry of compound films \cite{Kim2007, Sainty1984}.

The applied plasma in the CVD process enhances film uniformity and step coverage while improving adhesion and control over film growth. In PECVD, the precursors do not require high temperatures for the chemical reaction, which allows utilizing a temperature below 200 $^{\circ}$C with better uniformity, adhesion and deposition rate compared to magnetron sputtering \cite{Fenger2010}.

This paper aims to outline the typical dielectrics and related properties and the factors controlling their quality. We compare several silicon-based dielectric films grown under different conditions, which result in a substantial variation in nitrogen concentration and film stress. Then we present low-temperature loss measurements of the dielectric films using superconducting resonators. 

\section{Fabrication of Dielectrics and Test Devices}
The dielectric films were grown using several methods, including conventional MS and IBAS (Angstrom Engineering EvoVac), and PECVD (Oxford Plasmalab100), see Table 1 with the deposition parameters. We controlled film quality and stoichiometry by adjusting the applied power, gas mixture, and deposition pressure. The change of reactive gas flow from 0 to 20~sccm shifts the target condition from metallic to poisoned, thus different nitrogen gas flow allowed us to obtain from silicon to (Si-rich) nitride to silicon nitride films. We achieved high thickness uniformity at +/-2\% over a diameter of a 6-inch wafer. DC-power density for IBAS was kept at 20~$\mathrm{W/in^{2}}$ with max power of ion source (Kaufman and Robinson RF2100ICP) 1~kW. All IBAS depositions were performed at room temperature (RT). 

\begin{table*}[!t]
\caption{Dielectrics deposition parameters and properties\label{tab:table1}}
\centering
\begin{tabular}{|c|c|c|c|c|c|}
\hline
Material & Method & Deposition parameters & Ref. index & Stress & Tan delta \\
\hline
SiO2 & MS & 250$^{\circ}$C 4h, 4.6sccm $\mathrm{O_2}$ 25sccm Ar & 1.46 & -314 MPa & 1x$10^{-3}$ \\
\hline
SiO2 & IBAS & RT 1h, 8 sccm $\mathrm{O_2}$, 20sccm Ar & 1.58 & -311 MPa & 8x$10^{-4}$ \\
\hline
a-Si & IBAS & RT 1h, 20 sccm Ar & 4.5 & -705 MPa & 5x$10^{-4}$ \\
\hline
(Si-rich)Nx & IBAS & RT 1h, 6 sccm $\mathrm{N_2}$, 20sccm Ar & 2.4 & -1575 MPa & 8x$10^{-4}$ \\
\hline
SiNx & IBAS & RT 1h, 17 sccm $\mathrm{N_2}$, 20sccm Ar & 1.98 & -1702 MPa & $10^{-3}$ \\
\hline
(Si-rich)Nx & PECVD & 100$^{\circ}$C 1h, 33 sccm $\mathrm{SiH_4}$, 24 sccm $\mathrm{N_2}$ & 2.38 & -792 MPa & 2x$10^{-4}$ \\
\hline
SiNx & PECVD & 100$^{\circ}$C 1h, 30 sccm $\mathrm{SiH_4}$, 27 sccm $\mathrm{N_2}$ & 2.4 &  & 1.1x$10^{-3}$ \\
\hline
\end{tabular}
\end{table*}

\begin{figure}[!ht]
\centering
\includegraphics[width=0.95\linewidth]{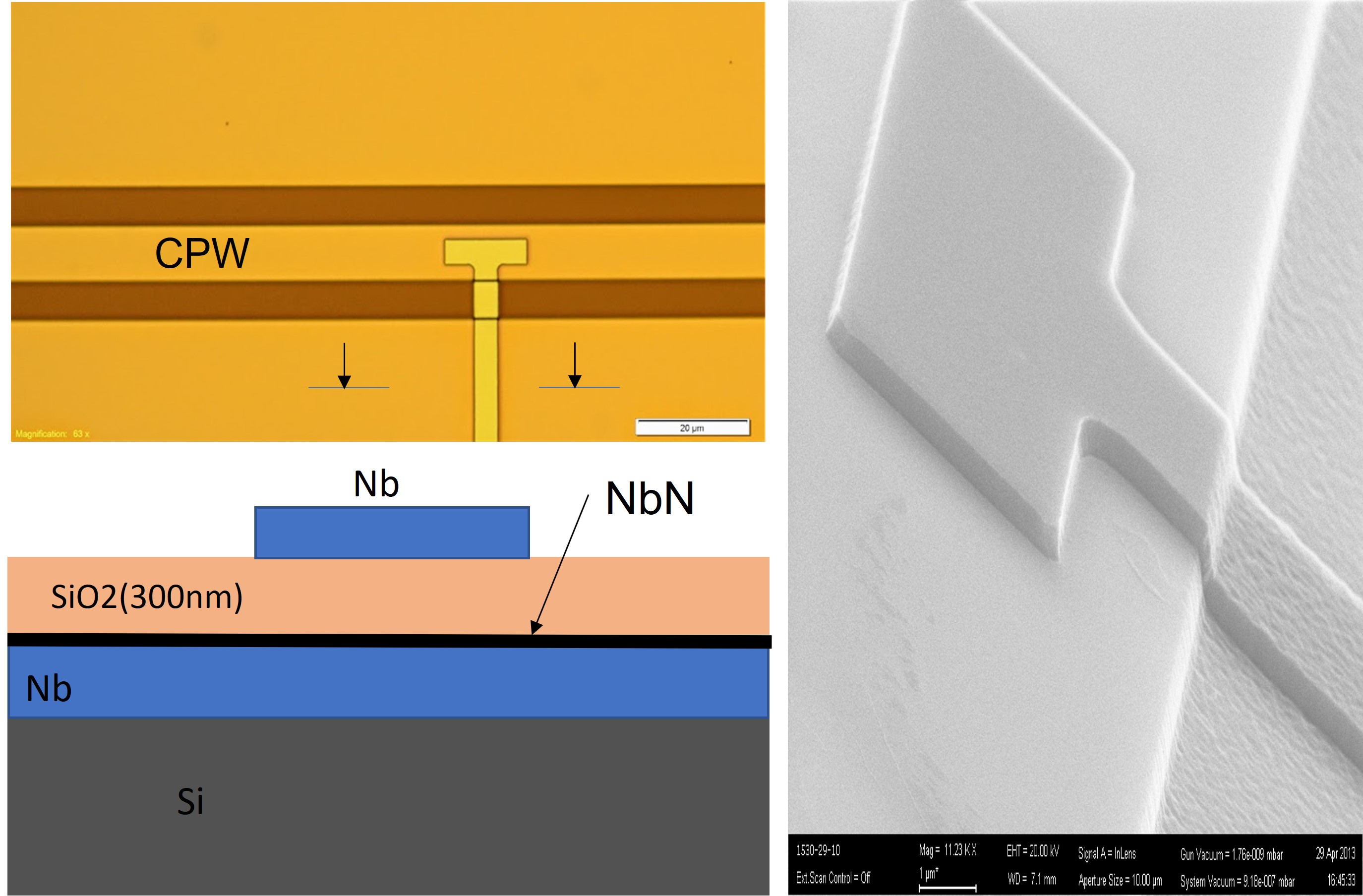}
\caption{Top left: microscope image of the microstrip resonator structure. A coplanar waveguide (CPW) feedline capacitively couples to an Nb microstrip resonator. Bottom left: schematics of the Nb microstrip resonator cross-section. The bottom layer of the Nb ground plane is deposited on a silicon wafer. We deposit the dielectric material in the middle and the Nb microstrip line on top of the dielectrics. Right: SEM image of the structure where Nb microstrip capacitively coupled to a CPW transmission line. }
\label{fig:fab}
\end{figure}

We designed and made half-wavelength microstrip resonators capacitively coupled to a coplanar waveguide (CPW) for loss measurements at GHz level (Fig. \ref{fig:fab}). Resonator length ranges from 7~mm to 8.5~mm. Fundamental resonant frequency is ~ 5 to 10 GHz depending on specific material parameters. More details on resonator design can be found in \cite{Chang2015}. We patterned test structures on a six-inch high-resistivity Si wafer with a three-layer design consisting of a 170~nm thick Nb ground plane with a 30~nm NbN cap, a 300~nm thick dielectric spacer, and a 400~nm thick Nb transmission line. The superconducting Nb films were deposited in an argon plasma at room temperature in a confocal DC magnetron sputtering system. The NbN capping layer was deposited following the bottom layer in-situ in the $\mathrm{Ar-N_2}$ gas, which protects the Nb layer underneath from oxidation during further processing before dielectric deposition. We rotate the substrate holder and adjust the tilt of the sputtering gun to provide high uniformity in film thickness. The base pressure was about ~$1\times10^{-8}$~Torr. 

Preliminarily measured superconducting transition temperatures of 100~nm Nb and NbN films showed  $\mathrm{T_c}$ 9.2~K and 14.5~K, respectively. Films were deposited in the same system on Si substrates. Test NbN films were sputtered onto the top of a 10~nm thick Nb seed layer.  Change of the nitrogen flow allowed variation of the NbN transition temperature from 10.5~K to 14.5~K.

Each layer of the microstrip resonator device was patterned with non-contact lithography using a maskless aligner (Heidelberg MLA 150) at 405~nm laser wavelength and UV-sensitive photoresist. The first two layers were fabricated by inductively coupled plasma (ICP) reactive ion etch. Nb/NbN was etched at 600~W with reactive gasses of $\mathrm{CHF_3}$ and $\mathrm{SF_6}$, while the dielectric layer was etched at a higher power of 1200~W with reactive gasses of $\mathrm{CHF_3}$ and $\mathrm{Ar}$. Last, the top layer microstrip was done by lift-off with double-layer resist to protect previous structures underneath and ensure smooth edges. 

\section{Film Characterization}
Visual inspection of fabricated structures provides vital feedback on possible defects that can cause resistive and dielectric losses. The structural features and morphological properties of patterned films were analyzed using optical and scanning electron microscopy. Figure \ref{fig:fab} shows the structure of the microstrip resonator used to measure the dielectric loss and the optical and SEM images of zoomed-in resonator segments. The Nb microstrip resonator is capacitively coupled to the Nb/NbN CPW transmission line separated by the dielectric layer ($\mathrm{SiO_x}$ for the sample shown) in the middle.   

\begin{figure}[!ht]
\centering
\includegraphics[width=0.95\linewidth]{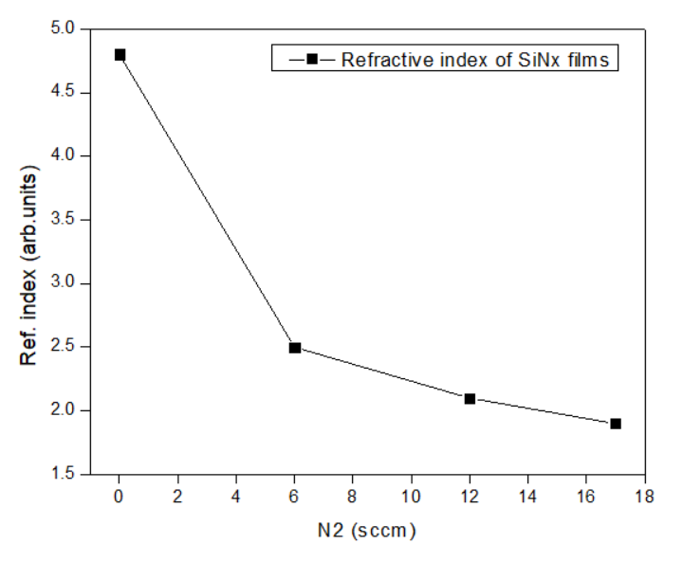}
\caption{Correlation between optical refraction index and nitrogen flow rate during a Si pulsed sputtering with an Ar flow of 20~sccm.}
\label{fig:ref_index}
\end{figure}

We characterized the dielectrics by measuring the dielectric films' optical properties, intrinsic stress, and surface morphology. Other researches confirm that the film's refractive index depends on the sputtering conditions, indicating a correlation with film stoichiometry and microstructure, and could be looked as film level of packaging and contamination doping. \cite{Xu2003}. The refractive index of dielectrics was characterized at room temperature using the spectrophotometer (Filmetrics FV20) operated at optical wavelengths of 632.8~nm. Measured parameters were correlated with film thickness and previos ellipsometer data. We found that the refractive index of $\mathrm{SiN_x}$ can change with nitrogen flow for the IBAS process, shown in Figure \ref{fig:ref_index}. The optical refraction index allows us to monitor and control the stoichiometry and degree of crystallinity of the dielectric films. We can also compare the dielectric constant with literature data for materials such as amorphous silicon and crystalline $\mathrm{Si_3N_4}$ 
 \cite{Ng2015}. 

\begin{figure}[!ht]
\centering
\includegraphics[width=0.95\linewidth]{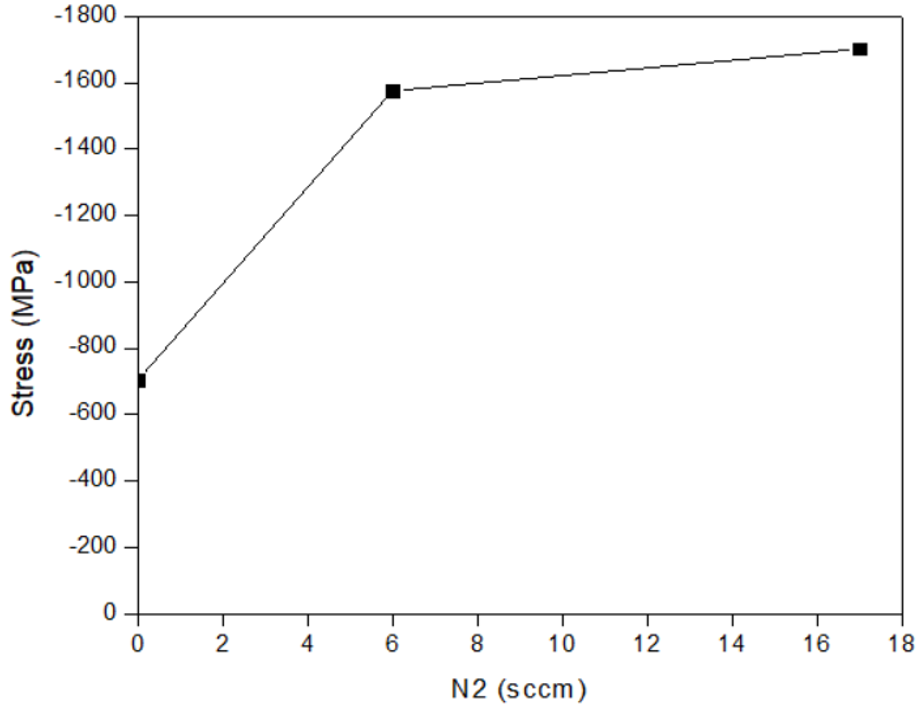}
\caption{Intrinsic stress of $\mathrm{SiN_x}$ films with different $\mathrm{N_2}$ concentrations.}
\label{fig:stress}
\end{figure}

The residual stress of the dielectric film can show the stability of the intrinsic structural bonds. According to Thornton model deposition parameters can highly impact the structure and stress of the growing film \cite{Thornton1986}. We measured the film’s stress using Multi-beam Optical Sensor (MOS) with UltraScan system, by laser array scanning and calculating the change of the wafer surface curvature before and after deposition. We can estimate how the deposition process impacts internal stress value from the stress results. The stress measurements of silicon-rich nitride by IBAS and PECVD differ by a factor of 2. We then changed the reactive gas flow change in IBAS films and observed a clear shift in the stress from highly to slightly compressive with pressures from -1702~MPa to -311~MPa (Fig. \ref{fig:stress}).  

\begin{figure}[!ht]
\centering
\includegraphics[width=0.95\linewidth]{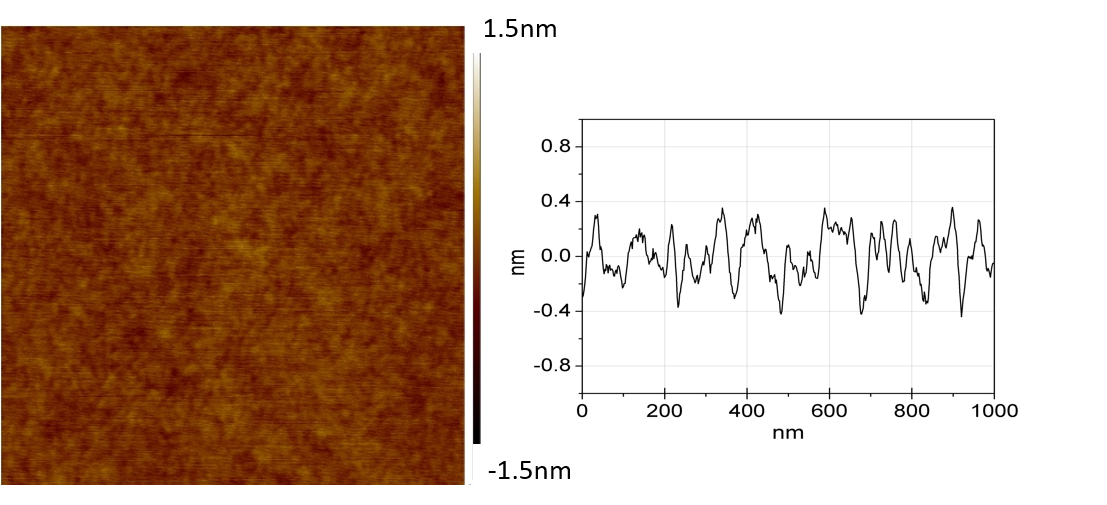}
\includegraphics[width=0.95\linewidth]{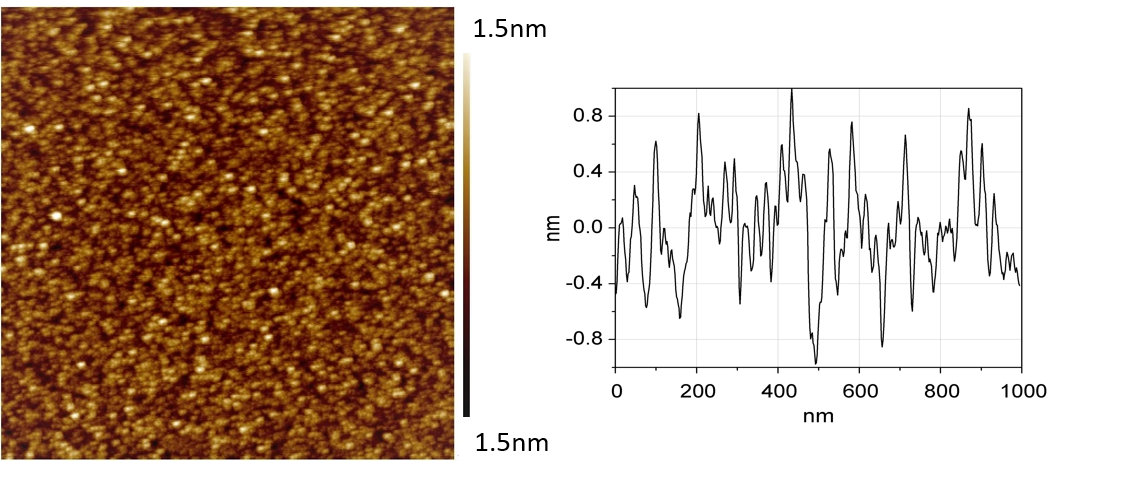}
\caption{AFM top surface images and roughness plots of the 300~nm-thick $\mathrm{SiN_x}$ (top) and $\mathrm{SiO_2}$ (bottom) films obtained by ion beam assisted sputtering.}
\label{fig:afm}
\end{figure}

Using atomic force microscopy (AFM), we can learn about the grain formation process and possible surface defects. Figure \ref{fig:afm} shows the top surface image and roughness profile of 300~nm-thick $\mathrm{SiO_2}$ and $\mathrm{SiN_x}$ films. $\mathrm{SiN_x}$ film scan shows a smoother surface with measured roughness of 0.4~nm. The average RMS roughnesses of all obtained dielectrics vary from 1.2 to 0.4~nm.  

\begin{figure}[!ht]
\centering
\includegraphics[width=0.95\linewidth]{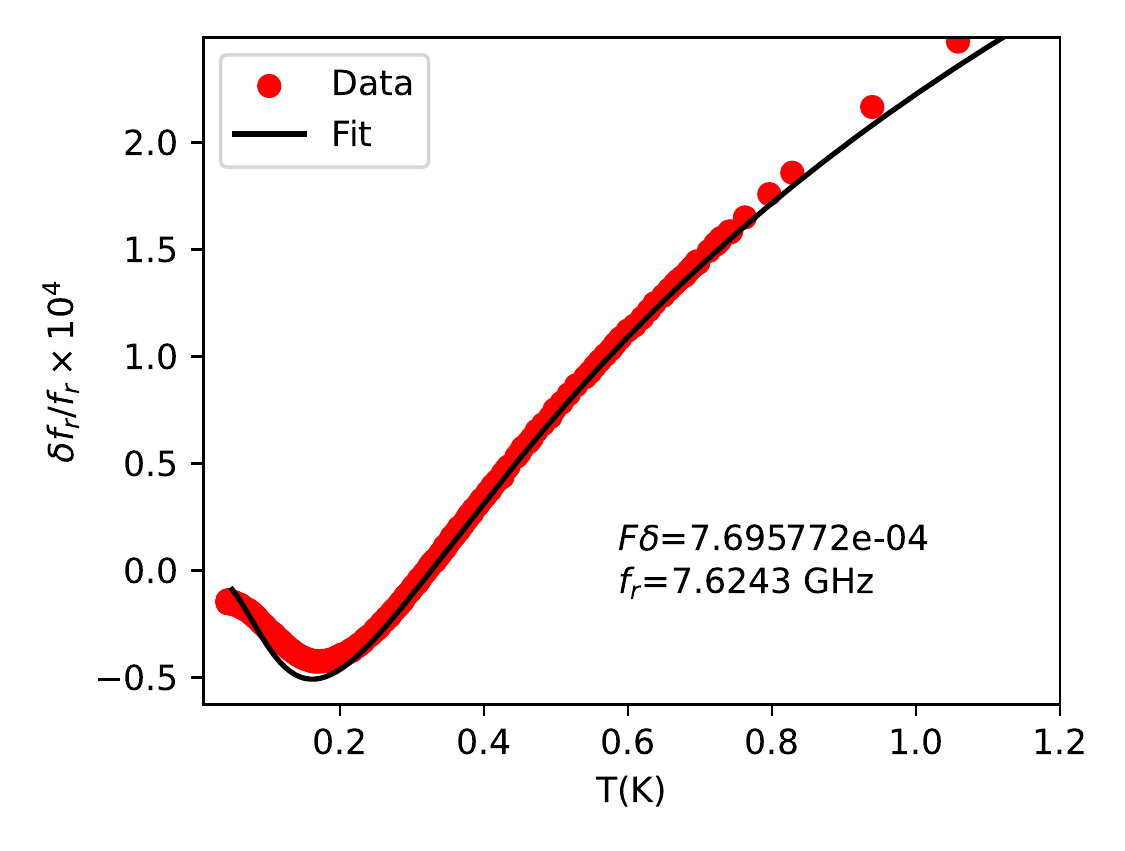}
\caption{A sample measurement of resonant frequency vs. operating temperature for a SiO$_2$ microstrip resonator at around 5~GHz. We fit the data into a TLS model to extract the loss of the dielectric material. }
\label{fig:sio2_loss}
\end{figure}

We studied the resonator loss at $\sim 4-10$~GHz frequencies by measuring the transmission of the probe signal as a function of frequency and temperature. A sample measurement of resonant frequency shift as a function of temperature for one IBAS SiO$_2$ sample is in Figure \ref{fig:sio2_loss}. The black line is the two-level system (TLS) model \cite{gao2008physics} fit to the measured data below 1.2~K. For the sample SiO$_2$ resonator, $f_r(0)=7.62$~GHz and $F\delta=7.7\times 10^{-4}$, where $f_r(0)$ is the resonant frequency, $\delta$ is the dielectric loss, and $F$ is the energy filling factor, which is $\sim 1$ for microstrip resonators. We performed similar measurements for the dielectric samples listed in Table \ref{tab:table1} and reported the loss numbers there. The loss tangent values were calculated from fits to frequency vs. temperature data taken over temperatures ranging from 15~mK to 1~K at frequencies of 5-8~GHz (see Fig. \ref{fig:sio2_loss} as an example). The dielectrics will be used for millimeter-wavelength (mm-wave) detectors, so it is also critical to measure their loss at mm-wave. We have finished the initial design, setup, and measurement of mm-wave dielectric loss and demonstrated a factor of two higher loss of $4\times 10^{-3}$ at 150~GHz for CVD SiN$_x$ compared to $2\times 10^{-3}$ at $\sim$1~GHz \cite{Pan2022}. Next, we plan to validate the dielectric performance at mm-wave.

\section{Conclusions}
In this paper, we summarize progress in developing fabrication techniques for superconducting microstrip structures. We discuss different approaches for high-quality dielectric material deposition with varying processing temperatures. We characterized the film quality using a variety of measurement technics, including refraction index characterization, surface quality imaging, and cryogenic loss measurements at ~GHz frequencies. Silicon-rich $\mathrm{SiN_x}$ deposited at 100$^{\circ}$C by PECVD showed low loss at $10^{-4}$ level and good surface-structural properties. One other promising material is $\mathrm{SiN_x}$ deposited at room temperature by IBAS with loss at $10^{-3}$ level. Looking ahead, we will improve quality control and reduce the loss further in sputtered dielectric films. We plan to extend our study to PECVD-grown Si-rich nitride with possible dielectric hydrogenation to terminate dangling bonds and defects.

\section{Acknowledgements}
Work at Argonne, including use of the Center for Nanoscale Materials (CNM), an Office of Science user facility, was supported by the U.S. Department of Energy, Office of Science, Office of Basic Energy Sciences, Materials Sciences and Engineering Division and Office of High Energy Physics, under Contract No. DE-AC02-06CH11357. 

\bibliographystyle{IEEEtran}
\bibliography{reference.bib}

\begin{thebibliography}{10}
\providecommand{\url}[1]{#1}
\csname url@samestyle\endcsname
\providecommand{\newblock}{\relax}
\providecommand{\bibinfo}[2]{#2}
\providecommand{\BIBentrySTDinterwordspacing}{\spaceskip=0pt\relax}
\providecommand{\BIBentryALTinterwordstretchfactor}{4}
\providecommand{\BIBentryALTinterwordspacing}{\spaceskip=\fontdimen2\font plus
\BIBentryALTinterwordstretchfactor\fontdimen3\font minus
  \fontdimen4\font\relax}
\providecommand{\BIBforeignlanguage}[2]{{%
\expandafter\ifx\csname l@#1\endcsname\relax
\typeout{** WARNING: IEEEtran.bst: No hyphenation pattern has been}%
\typeout{** loaded for the language `#1'. Using the pattern for}%
\typeout{** the default language instead.}%
\else
\language=\csname l@#1\endcsname
\fi
#2}}
\providecommand{\BIBdecl}{\relax}
\BIBdecl

\bibitem{Buijtendorp2020}
B.~T. Buijtendorp, J.~Bueno, D.~J. Thoen, V.~Murugesan, P.~Sberna, J.~J.~A.
  Baselmans, S.~Vollebregt, and A.~Endo, ``Characterization of low-loss
  hydrogenated amorphous silicon films for superconducting resonators,''
  \emph{Journal of Astronomical Telescopes, Instruments, and Systems}, vol.~8,
  pp. 028\,006 -- 028\,006, 2020.

\bibitem{Paik2010}
H.~Paik and K.~D. Osborn, ``Reducing quantum-regime dielectric loss of silicon
  nitride for superconducting quantum circuits,'' \emph{Applied Physics
  Letters}, vol.~96, p. 072505, 2010.

\bibitem{Xu2003}
\BIBentryALTinterwordspacing
G.~Xu, P.~Jin, M.~Tazawa, and K.~Yoshimura, ``Optical investigation of silicon
  nitride thin films deposited by r.f. magnetron sputtering,'' \emph{Thin Solid
  Films}, vol. 425, no.~1, pp. 196--202, 2003. [Online]. Available:
  \url{https://www.sciencedirect.com/science/article/pii/S0040609002010891}
\BIBentrySTDinterwordspacing

\bibitem{Tang2017}
\BIBentryALTinterwordspacing
C.-J. Tang, C.-C. Jaing, C.-L. Tien, W.-C. Sun, and S.-C. Lin, ``Optical,
  structural, and mechanical properties of silicon oxynitride films prepared by
  pulsed magnetron sputtering,'' \emph{Appl. Opt.}, vol.~56, no.~4, pp.
  C168--C174, Feb 2017. [Online]. Available:
  \url{https://opg.optica.org/ao/abstract.cfm?URI=ao-56-4-C168}
\BIBentrySTDinterwordspacing

\bibitem{Cerny1998}
\BIBentryALTinterwordspacing
F.~Cerný, J.~Suchánek, and V.~Hnatowicz, ``Ibad sinx coating: preparation and
  tribological properties,'' \emph{Thin Solid Films}, vol. 317, no.~1, pp.
  490--492, 1998. [Online]. Available:
  \url{https://www.sciencedirect.com/science/article/pii/S0040609097005701}
\BIBentrySTDinterwordspacing

\bibitem{Freer1990}
\BIBentryALTinterwordspacing
R.~Freer and I.~O. Owate, \emph{The Dielectric Properties of Nitrides}.\hskip
  1em plus 0.5em minus 0.4em\relax Dordrecht: Springer Netherlands, 1990, pp.
  639--651. [Online]. Available:
  \url{https://doi.org/10.1007/978-94-009-2101-6_36}
\BIBentrySTDinterwordspacing

\bibitem{Chang2015}
C.~L. Chang, P.~A.~R. Ade, Z.~Ahmed, S.~W. Allen, K.~Arnold, J.~E. Austermann,
  A.~N. Bender, L.~E. Bleem, B.~A. Benson, J.~E. Carlstrom, H.~M. Cho, S.~T.
  Ciocys, J.~F. Cliche, T.~M. Crawford, A.~Cukierman, J.~Ding, T.~de~Haan,
  M.~A. Dobbs, D.~Dutcher, W.~Everett, A.~Gilbert, N.~W. Halverson, D.~Hanson,
  N.~L. Harrington, K.~Hattori, J.~W. Henning, G.~C. Hilton, G.~P. Holder,
  W.~L. Holzapfel, J.~Hubmayr, K.~D. Irwin, R.~Keisler, L.~Knox, D.~Kubik,
  C.~L. Kuo, A.~T. Lee, E.~M. Leitch, D.~Li, M.~McDonald, S.~S. Meyer,
  J.~Montgomery, M.~Myers, T.~Natoli, H.~Nguyen, V.~Novosad, S.~Padin, Z.~Pan,
  J.~Pearson, C.~Posada~Arbelaez, C.~L. Reichardt, J.~E. Ruhl, B.~R.
  Saliwanchik, G.~Simard, G.~Smecher, J.~T. Sayre, E.~Shirokoff, A.~A. Stark,
  K.~Story, A.~Suzuki, K.~L. Thompson, C.~Tucker, K.~Vanderlinde, J.~D. Vieira,
  A.~Vikhlinin, G.~Wang, V.~Yefremenko, and K.~W. Yoon, ``Low loss
  superconducting microstrip development at argonne national lab,'' \emph{IEEE
  Transactions on Applied Superconductivity}, vol.~25, no.~3, pp. 1--5, 2015.

\bibitem{Li2013}
D.~Li, J.~Gao, J.~E. Austermann, J.~A. Beall, D.~Becker, H.-M. Cho, A.~E. Fox,
  N.~Halverson, J.~Henning, G.~C. Hilton, J.~Hubmayr, K.~D. Irwin,
  J.~Van~Lanen, J.~Nibarger, and M.~Niemack, ``Improvements in silicon oxide
  dielectric loss for superconducting microwave detector circuits,'' \emph{IEEE
  Transactions on Applied Superconductivity}, vol.~23, no.~3, pp.
  1\,501\,204--1\,501\,204, 2013.

\bibitem{oconnel2008}
A.~D. O’Connell, M.~Ansmann, R.~C. Bialczak, M.~Hofheinz, N.~Katz, E.~Lucero,
  C.~McKenney, M.~Neeley, H.~Wang, E.~M. Weig, A.~N. Cleland, and J.~M.
  Martinis, ``Microwave dielectric loss at single photon energies and
  millikelvin temperatures,'' \emph{Applied Physics Letters}, vol.~92, no.~11,
  p. 112903, 2008.

\bibitem{Kim2007}
C.~Kim, S.~Jo, J.~Kim, S.~Ryu, J.~H. Noh, H.~Baik, S.~Lee, and Y.~Kim, ``Low
  leakage current gate dielectrics prepared by ion beam assisted deposition for
  organic thin film transistors,'' \emph{Journal of Applied Physics}, vol. 102,
  pp. 126\,101--126\,101, 12 2007.

\bibitem{Sainty1984}
\BIBentryALTinterwordspacing
W.~G. Sainty, R.~P. Netterfield, and P.~J. Martin, ``Protective dielectric
  coatings produced by ion-assisted deposition,'' \emph{Appl. Opt.}, vol.~23,
  no.~7, pp. 1116--1119, Apr 1984. [Online]. Available:
  \url{https://opg.optica.org/ao/abstract.cfm?URI=ao-23-7-1116}
\BIBentrySTDinterwordspacing

\bibitem{Fenger2010}
G.~Fenger, ``Development of plasma enhanced chemical vapor deposition (pecvd)
  gate dielectrics for tft applications,'' 2010.

\bibitem{Ng2015}
D.~Ng, Q.~Wang, T.~Wang, S.~Ng, Y.~Toh, K.~P. Lim, Y.~Yang, and D.~Tan,
  ``Exploring high refractive index silicon rich nitride films by low
  temperature inductively-coupled plasma chemical vapor deposition and
  applications for integrated waveguides,'' \emph{ACS applied materials and
  interfaces}, vol.~7, 09 2015.

\bibitem{Thornton1986}
J.~A. Thornton, ``The microstructure of sputter-deposited coatings,''
  \emph{Journal of Vacuum Science and Technology}, vol.~4, pp. 3059--3065,
  1986.

\bibitem{gao2008physics}
J.~Gao, \emph{The physics of superconducting microwave resonators}.\hskip 1em
  plus 0.5em minus 0.4em\relax California Institute of Technology, 2008.

\bibitem{Pan2022}
Z.~Pan, P.~Barry, T.~Cecil, C.~Albert, A.~Bender, C.~Chang, R.~Gualtieri,
  J.~Hood, J.~Li, J.~Zhang, M.~Lisovenko, V.~Novosad, G.~Wang, and
  V.~Yefremenko, ``Measurements of dielectric material loss from millimeter to
  centimeter wavelengths,'' \emph{Presented at ASC 2022, Honolulu, Hawaii, USA,
  October 2022, Paper number ASC2022-3EPo2D-02}, p. N. A., 2022.

\end{thebibliography}

\end{document}